\documentclass[twocolumn,showpacs,preprintnumbers,amsmath,amssymb]{revtex4}

\usepackage{graphicx}
\usepackage{dcolumn}
\usepackage{bm}

\begin{document}

\title{Thermodynamics of ferromagnetic superconductors with spin-triplet electron pairing}

\author{Diana V. Shopova}

\author{Dimo I. Uzunov$^1$}

\altaffiliation[]{Corresponding author. Electronic address:
d.i.uzunov@gmail.com}

\affiliation{CP Laboratory, Institute of Solid State Physics,
Bulgarian Academy of Sciences, BG-1784 Sofia, Bulgaria,}

\affiliation{$^{1}$ Department of Physics and Astronomy,
University of Western Ontario, London, Ontario N6A 3K7, Canada.}

\date{\today}

\begin{abstract}
We present a general thermodynamic theory that describes  phases
and phase transitions of  ferromagnetic superconductors with
spin-triplet electron Cooper pairing. The theory is based on
extended Ginzburg-Landau expansion in powers of superconducting
and ferromagnetic order parameters. We propose a simple form for
the dependence of  theory parameters  on the pressure that allows
correct theoretical outline of the temperature-pressure phase
diagram for which at low temperatures a  stable phase of
coexistence of $p$-wave superconductivity and itinerant
ferromagnetism  appears. We demonstrate that the theory is in an
agreement with the experimental data for some intermetallic
compounds that are experimentally proven to be itinerant
ferromagnetic exhibiting spin-triplet superconductivity.  Some
basic features of quantum phase transitions in such systems are
explained and clarified. We propose to group the spin-triplet
ferromagnetic superconductors in two different types of
thermodynamic behavior, on the basis of quantitative criterion
deduced from the present theory and the analysis of experimental
data.
\end{abstract}



\maketitle

\section{\label{sec:level1}Introduction}

The spin-triplet  or $p$-wave pairing allows parallel spin
orientation of the fermion Cooper pairs in superfluid $^3$He and
unconventional superconductors~\cite{Vollhardt:1990}. For this
reason the resulting unconventional superconductivity is robust
with respect to effects of external magnetic field and spontaneous
ferromagnetic order, so it may coexist with the latter. This
general argument implies that there could be metallic compounds
and alloys, for which the  coexistence of spin-triplet
superconductivity and ferromagnetism may be observed.

Particularly, both superconductivity and itinerant ferromagnetic
orders can be created by the same band electrons in the metal,
which means that spin-1 electron Cooper pairs participate in the formation
of the itinerant ferromagnetic order. Moreover, under certain
conditions the superconductivity is enhanced rather than
depressed by the uniform ferromagnetic order that  can generate
it, even in cases when the superconductivity does not appear in a pure form
as a net result of indirect electron-electron coupling.

 The   coexistence
of superconductivity and ferromagnetism as a result of collective
behavior of  $f$-band electrons has been found experimentally for
some Uranium-based intermetallic compounds as,
UGe$_2$~\cite{Saxena:2000,Huxley:2001, Tateiwa:2001, Harada:2007},
URhGe~\cite{Aoki:2001, Hardy1:2005, Hardy2:2005},
UCoGe~\cite{Huy:2007, Huy:2008}, and UIr~\cite{Akazawa:2005,
Kobayashi:2006}. At low temperature $(T \sim 1~\mbox{K})$ all
these compounds exhibit thermodynamically stable phase of
coexistence of spin-triplet superconductivity and itinerant
($f$-band) electron ferromagnetism (in short, FS phase). In
UGe$_2$ and UIr the FS phase appears at high pressure $(P \sim
1~\mbox{GPa})$ whereas in URhGe and UCoGe,  the coexistence  phase
persists up to ambient pressure $(10^5 \mbox{Pa} \equiv 1
\mbox{bar})$.

Experiments, carried out in ZrZn$_2$ ~\cite{Pfleiderer:2001},
also indicated the appearance of  FS phase   at $ T < 1$ K in a
wide range of pressures ($0<P \sim 21~\mbox{kbar}$). In Zr-based
compounds the ferromagnetism and the $p$-wave superconductivity
occur as a result of the collective behavior of the $d$-band
electrons. Later experimental results~\cite{Yelland1:2005,
Yelland2:2005} had imposed the conclusion that  bulk
superconductivity is lacking in ZrZn$_2$,  but the occurrence of a
surface FS phase at surfaces with higher Zr content than that in
ZrZn$_2$ has been reliably demonstrated. Thus the problem for the
coexistence of bulk superconductivity with ferromagnetism in
ZrZn$_2$ is still unresolved. This raises the question whether the
FS phase in ZrZn$_2$ should be studied by  surface thermodynamics
methods, or it can be investigated by considering that  bulk and
surface thermodynamic phenomena can be  treated on the same
footing. Taking in account the mentioned experimental results for
ZrZn$_2$ and their interpretation by the
experimentalists~\cite{Pfleiderer:2001,Yelland1:2005,
Yelland2:2005} we assume that the unified thermodynamic approach
can be applied, although some specific properties of the FS phase
in ZrZn$_2$, if its surface nature is confirmed, will certainly
need special
study by surface thermodynamics. \\
Here we will investigate the itinerant ferromagnetism and
superconductivity of U- and Zr-based intermetallic compounds
within the same general thermodynamic approach. Arguments
supporting our point of view are given in several preceding
investigations. We should mention that the spin-triplet
superconductivity occurs not only in bulk materials, but also in
quasi-two-dimensional (2D) systems
 - thin films and surfaces and quasi-1D wires (see, e.g.,
Refs.~\cite{Bolesh:2005}). In ZrZn$_2$ both
ferromagnetic and superconducting orders  vanish at the
same critical pressure $P_c$, a fact  implying that the respective order
parameter fields strongly depend on each other and should be
studied on the same thermodynamic
basis~\cite{Nevidomskyy:2005}.

The general thermodynamic treatment does not necessarily specify
the system spatial dimensionality $D$: $D=3$ describes the bulk
properties, and $D=2$ --  very thin films and mono-atomic layers.
Within  Landau theory of phase transitions (see, e.g.,
Ref.~\cite{Uzunov:1993}), the system dimensionality can be
distinguished by the values of the Landau parameters. Here we
specify the values of these parameters from the experimental data
for spin-triplet ferromagnetic superconductors. When the Landau
parameters are obtained from microscopic theories, their values
depend on the dimension $D$, only if the respective theory takes
into account relevant fluctuation modes of order parameter fields,
including long-scale fluctuation modes. We are not aware of well
developed theories of this type which may figure out the complex
behavior of the mentioned systems. Even in simple theories of band
electron magnetism, the Landau parameters are very complex
functions of the density of states at the Fermi level and related
microscopic parameters. Such complexity does not allow  direct
comparison between the results from microscopic theory  and the
experimental data. To make a progress in this situation we assume
that the material parameters of our theory are  loosely defined
and may have values, corresponding to various approximate
microscopic theories, as mean-field approximation,
spin-fluctuation theory, etc.

For all  compounds, cited above,  the FS phase occurs only in the
ferromagnetic phase domain of the $T-P$ diagram. Particularly at
equilibrium, and for  given $P$, the temperature $T_{F}(P)$ of the
normal-to-ferromagnetic phase (or N-FM) transition is never lower
than the temperature $T_{FS}(P)$ of the ferromagnetic-to-FS phase
transition (FM-FS transition). This confirms  the point of view
that the superconductivity in these compounds is triggered by the
spontaneous magnetization $\mbox{\boldmath$M$}$, in analogy with
the well-known triggering of the superfluid phase A$_1$ in $^3$He
at mK temperatures by the external magnetic field
$\mbox{\boldmath$H$}$. Such ``helium analogy" has been used in
some theoretical studies (see, e.g., Ref.~\cite{Machida:2001,
Walker:2002}), where Ginzburg-Landau (GL) free energy terms,
describing the FS phase were derived by symmetry group arguments.
The non-unitary state, with a non-zero value of the Cooper pair
magnetic moment, known from the theory of unconventional
superconductors and superfluidity in $^3$He~\cite{Vollhardt:1990},
has been suggested firstly in Ref.~\cite{Machida:2001}, and later
confirmed in other studies~\cite{Hardy1:2005, Walker:2002};
recently, the same topic was comprehensively discussed
in~\cite{Linder:2007}.

For the spin-triplet ferromagnetic superconductors the trigger
mechanism was recently examined in detail~\cite{Shopova:2005,
Shopova:2006}. The system main properties are specified by the
term in the GL expansion of  form $\mbox{\boldmath$M$}
|\mbox{\boldmath$\psi$}|^2$, which represents the interaction of
$\mbox{\boldmath$M$} = \{M_j; j=1,2,3\}$ with the complex
superconducting vector field $\mbox{\boldmath$\psi$} =\{\psi_j\}$.
Particularly, this term is responsible for the appearance of
$\mbox{\boldmath$\psi$} \neq 0$ for certain $T$ and $P$ values. A
similar trigger mechanism is familiar in the context of improper
ferroelectrics~\cite{Cowley:1980}.

A crucial feature of these  systems is the nonzero magnetic moment
of the spin-triplet Cooper pairs. As mentioned above, the
microscopic theory of magnetism and superconductivity in non-Fermi
liquids of strongly interacting heavy electrons ($f$ and $d$ band
electrons) is either too complex or insufficiently developed to
describe the complicated behavior in itinerant ferromagnetic
compounds. Several authors (see~\cite{Machida:2001, Walker:2002,
Shopova:2005, Linder:2007}) have explored the phenomenological
description by a self-consistent mean field theory, and we will
essentially use the thermodynamic results, in particular, the
analysis in Refs.~\cite{Shopova:2005, Shopova:2006}. Mean-field
microscopic theory of spin-mediated pairing leading to the
mentioned non-unitary superconductivity state has been developed
in Ref.~\cite{Nevidomskyy:2005} that  is in conformity with the
phenomenological description that we have done.

In this paper, we present general thermodynamic treatment of
systems with itinerant ferromagnetic order and superconductivity
due to spin-triplet Cooper pairing of the same band electrons,
which are responsible for the spontaneous magnetic moment. We
outline their $T-P$ phase diagrams and demonstrate two contrasting
types of thermodynamic behavior. The present phenomenological
approach includes both mean-field and spin-fluctuation theory
(SFT), as the arguments in Ref.~\cite{Yamada:1993}. We propose a
simple, yet comprehensive, modelling of  $P$ dependence of the
free energy parameters, resulting in a very good compliance of
our theoretical predictions for the shape the $T-P$ phase diagrams
with the experimental data (for some preliminary results, see
Ref.~\cite{Cottam:2008}.

The theoretical analysis is  done by the standard methods of
 phase transition theory~\cite{Uzunov:1993}. Treatment of
fluctuation effects and quantum correlations~\cite{Uzunov:1993,
Shopova:2003} is not included in this study. But the parameters of
the generalized GL free energy may be considered either in
mean-field approximation as here, or  as phenomenologically
re-normalized parameters which are affected by additional physical
phenomena, as for example, spin fluctuations.

We demonstrate with the help of  present theory that we can
outline different possible topologies for the $T-P$ phase diagram,
depending on the values of Landau parameters, derived from the
existing  experimental data. We show that for unconventional
(spin-triplet) ferromagnetic superconductors (FSs) there exist two
distinct types of behavior, which we denote as Zr-type (or,
alternatively, type I) and U-type (or, type II). This
classification of the FS, first mentioned in
Ref.~\cite{Cottam:2008}, is based on the reliable
interrelationship between a quantitative criterion derived by us
and the thermodynamic properties of the spin-triplet FSs (see Sec.
III.C and Sec. IV.D). Our approach can be also applied to URhGe,
UCoGe, and UIr. Our results shed light on the problems connected
with the order of the quantum phase transitions at ultra-low and
zero temperatures. They also raise the question for further
experimental investigations of the detailed structure of the phase
diagrams in the high-$P$/low-$T$ region.

In Sec. II we present the GL free energy of ferromagnetic
superconductors with spin-triplet Cooper pairing.
Here the arguments for the proposed simple pressure dependence of the
Landau  parameters are explained  and, as the comparison with
experimental data demonstrates, the chosen P-dependence  gives quite
satisfactory quantitative results. In more detailed studies one
may use more complex pressure dependence of  material
parameters. In Sec. III we present  our  theoretical results for the
various possible shapes of the phase diagram, multi-critical
points, quantum phase transitions. The theory predicts all types
of phase transition points and lines which are needed in the
description of real itinerant ferromagnets with spin-triplet
Cooper pairing.  In Sec. IV we compare our
theoretical results with the experimental data for some
intermetallic compounds. The comparison gives very good
quantitative agreement between experiment and theory for all known
experimental examples, except URhGe, where a
particular form of pressure dependence of the material parameters
should be suggested. On the basis of this analysis we propose a
possible classification of unconventional ferromagnetic
superconductors, showing on the basis of quantitative criterion
that they are of two types (Sec. IV.E). In Sec. V we summarize our
findings and some unresolved problems which are beyond the scope
of the present study.

\section{\label{sec:level1}Generalized Ginzburg-Landau free energy}

Following Ref.~\cite{Shopova:2005}, the free energy per unit
volume, $F/V =f(\mbox{\boldmath$\psi$},\mbox{\boldmath$M$})$, can
be written in the form

\begin{widetext}
\begin{equation}
\label{Eq1} f(\mbox{\boldmath$\psi$},\mbox{\boldmath$M$})=
  a_s|\mbox{\boldmath$\psi$}|^2 +\frac{b_s}{2}|\mbox{\boldmath$\psi$}|^4 +
  \frac{u_s}{2}|\mbox{\boldmath$\psi$}^2|^2 +
\frac{v_s}{2}\sum_{j=1}^{3}|\psi_j|^4 +
 a_f\mbox{\boldmath$M$}^2 +
 \frac{b_f}{2}\mbox{\boldmath$M$}^4 + i\gamma_0 \mbox{\boldmath$M$}
 \cdot (\mbox{\boldmath$\psi$}\times \mbox{\boldmath$\psi$}^*) + \delta
\mbox{\boldmath$M$}^2 |\mbox{\boldmath$\psi$}|^2.
\end{equation}
\end{widetext}

\noindent The material parameters satisfy $b_s
>0$, $b_f>0$; $a_s = \alpha_s(T-T_{s})$, and $a_f =
\alpha_{f}[T^n-T_{f}^n(P)]$, where $n=1$ gives the standard form
of $a_f$, and $n=2$ applies for SFT~\cite{Yamada:1993} and the
Stoner-Wohlfarth model~\cite{ Wohlfarth:1968}. The terms
proportional to $u_s$ and $v_s$ describe, respectively, the
anisotropy of the spin-triplet electron Cooper pairs and the
crystal anisotropy. Next, $\gamma_0 \sim J$ (with $J>0$, the
ferromagnetic exchange constant) and $\delta > 0$ are parameters
of the $\mbox{\boldmath$\psi$}$-$\mbox{\boldmath$M$}$ interaction
terms. Previous studies~\cite{Shopova:2005} have shown that the
anisotropy represented by the $u_s$ and $v_s$ terms in
Eq.~(\ref{Eq1}), slightly perturbs the size and shape of the
stability domains of the phases, while similar effects can be
achieved by varying the $b_s$ factor in the
$b_s|\mbox{\boldmath$\psi$}|^4$ term. For such reasons, in the
present analysis we ignore the anisotropy terms, setting $u_s =
v_s = 0$, and consider $b_s\equiv b >0$ as an effective parameter.
Then, without loss of generality, we may choose the
magnetization vector to have the form $\mbox{\boldmath$M$} =
(0,0,M)$.

According to the microscopic theory of band magnetism and
superconductivity, the macroscopic material parameters in
Eq.~(\ref{Eq1}) depend in a very complex way on the density of
states at the Fermi level and related microscopic
quantities~\cite{Misra:2008}. That is why we  can hardly use
the microscopic characteristics of these complex metallic
compounds in order to elucidate their thermodynamic properties, in
particular, in outlining their phase diagrams in some details.
However, some simple microscopic models give useful
results, for example, the zero temperature Stoner-type model
employed in Ref.~\cite{Sandeman:2003}.

We redefine for convenience the free energy,  Eq.~(\ref{Eq1}), in
a dimensionless form by $\tilde{f} = f/(b_f M_0^4)$, where $M_0 =
[\alpha_fT_{f0}^n /b_f]^{1/2} >0$ is the value of the
magnetization $M$ corresponding to the pure magnetic subsystem
$(\mbox{\boldmath$\psi$} \equiv 0)$ at $T=P=0$ and
$T_{f0}=T_f(0)$. Order parameters assume the scaling $m = M/M_0$
and $\mbox{\boldmath$\varphi$} = \mbox{\boldmath$\psi$}
/[(b_f/b)^{1/4}M_0]$ and in result, the free energy becomes

\begin{equation}
\label{Eq2} \tilde{f}= r\phi^2 + \frac{\phi^4}{2}+ tm^2
+\frac{m^4}{2} + 2\gamma m\phi_1\phi_2\mbox{sin}\theta +
\gamma_1m^2\phi^2,
\end{equation}

\noindent where $\phi_j =|\varphi_j|$, $(j=1,2,3)$, $\phi =
|\mbox{\boldmath$\varphi$}|$, and $\theta = (\theta_2 - \theta_1)$
is the phase angle between the complex $\varphi_1 = \phi_1
e^{i\theta_1}$ and $\varphi_2 = \phi_2 e^{\theta_2}$. Note, that
the third phase angle, $\theta_3$, corresponding to the third
complex field component $\varphi_3 =\phi_3e^{i\theta_3}$, does not
enter explicitly in the free energy density $\tilde{f}$, given by
Eq.~(\ref{Eq2}), which is a natural result of the continuous space
degeneration. The dimensionless parameters $t$, $r$, $\gamma$ and
$\gamma_1$ in Eq.~(\ref{Eq2}) are given by

\begin{equation}
\label{Eq3} t = \tilde{T}^n-\tilde{T}_f^n(P),\;\;\;\; r = \kappa
(\tilde{T}-\tilde{T}_s),
\end{equation}
where $\kappa = \alpha_sb_f^{1/2}/\alpha_fb^{1/2}T_{f0}^{n-1}$,
$\gamma = \gamma_0/ [\alpha_fT_{f0}^nb]^{1/2}$, and $\gamma_1 =
\delta/(bb_f)^{1/2}$. The reduced temperatures are: $\tilde{T} =
T/T_{f0}$, $\tilde{T}_f(P) = T_f(P)/T_{f0}$, and $\tilde{T}_s(P)=
T_s(P)/T_{f0}$.

Here we will outline the  simple assumptions for the
$P$-dependence of the $t$, $r$, $\gamma$, and $\gamma_1$
parameters in Eq.~(\ref{Eq2}). Particularly, we assume that only
$T_f$ has a significant $P$-dependence, described by

\begin{equation}
\label{Eq4} \tilde{T}_f(P) = (1 - \tilde{P})^{1/n},
\end{equation}
where $\tilde{P} = P/P_0$ and $P_0$ is a characteristic pressure
deduced later. In ZrZn$_2$ and UGe$_2$ the $P_0$ values are very
close to the critical pressure $P_c$, at which both the
ferromagnetic and superconducting orders vanish, but in other
systems this is not necessarily the case. As we will discuss, the
nonlinearity ($n=2$) of $T_f(P)$ in ZrZn$_2$ and UGe$_2$ is
relevant at relatively high $P$, at which the N-FM transition
temperature $T_F(P)$ may not coincide with $T_f(P)$. The form
(\ref{Eq3}) of the model function $\tilde{T}_f(P)$ is consistent
with preceding experimental and theoretical investigations of the
N-FM phase transition in ZrZn$_2$ and UGe$_2$ (see, e.g.,
Refs.~\cite{Tateiwa:2001, Walker:2002, Smith:1971}). Here we
consider only non-negative values of the pressure $P$; for effects
at $P<0$, see, e.g., Ref.~\cite{Kimura:2004}.

The model function~(\ref{Eq4}) is defined for $P \leq P_0$, in
particular, for the case of $n >1$, but we should have in mind
that, in fact, the thermodynamic analysis of Eq.~(\ref{Eq2})
includes the parameter $t$ rather than $T_f(P)$. This parameter is
given by

\begin{equation}
\label{Eq5} t(T,P) = \tilde{T}^n - 1 + \tilde{P},
\end{equation}
and is well defined for any $\tilde{P}$. This allows to consider
 pressures $P > P_0$ within the free
energy~(\ref{Eq2}).

The model function $\tilde{T}_f(P)$ can be naturally generalized
to $\tilde{T}_f(P) = (1-\tilde{P}^{\beta})^{1/\alpha}$ but the
present needs of interpretation of experimental data do not
require this; hereafter we use
Eq.~(\ref{Eq3}) which corresponds to $\beta = 1$ and $\alpha =
n$). Besides, other analytical forms of $\tilde{T}_f(\tilde{P})$
can also be tested in the free energy~(\ref{Eq2}), in particular,
expansion in powers of $\tilde{P}$, or, alternatively, in $(1 -
\tilde{P})$ which satisfy the conditions $\tilde{T}_f(0) = 1$ and
$\tilde{T}_f(1) = 0$. Note, that in URhGe the slope of
$T_{F}(P)\sim T_f(P)$ is positive from $P=0$ up to high
pressures~\cite{Hardy2:2005} and for this compound the form
(\ref{Eq3}) of $\tilde{T}_f(P)$ is inconvenient. Here we apply the
simplest form for the $P$-dependence, namely, Eqs.~(\ref{Eq4})~--~(\ref{Eq5}).

\section{Phase transitions and multicritical points}

\subsection{Phase of coexistence of ferromagnetism and superconductivity}

The simplified model~(\ref{Eq2}) is capable of describing the main
thermodynamic properties of spin-triplet ferromagnetic
superconductors. For $r >0$, i.e., $T > T_s$, there are three
stable phases: (i) the normal (N-) phase, given by $\phi = m = 0$
(stability conditions: $t\geq0$, $r \geq 0$); (ii) the pure
ferromagnetic phase (FM phase), given by $m = (-t)^{1/2}> 0$,
$\phi =0$, which exists for $t < 0$ and is stable, provided $r\geq
0$ and $r \geq (\gamma_1t + \gamma|t|^{1/2})$, and (iii) the
 phase of coexistence of ferromagnetic order and
superconductivity (FS phase), given by $\mbox{sin}\theta = \mp 1$,
$\phi_3 = 0$, $\phi_1=\phi_2= \phi/\sqrt{2}$, where

\begin{equation}
\label{Eq6} \phi^2 = \kappa (\tilde{T}_s-\tilde{T}) \pm \gamma m -
\gamma_1 m^2 \geq 0.
\end{equation}
The magnetization  $m$ satisfies the equation

\begin{equation}
\label{Eq7} c_3 m^3 \pm c_2 m^2 + c_1 m \pm c_0 = 0
\end{equation}
with coefficients $c_0 = \gamma\kappa(\tilde{T} - \tilde{T}_s)$,
\begin{equation}
\label{Eq8}  c_1 = 2\left[\tilde{T}^n + \kappa\gamma_1(\tilde{T}_s
-\tilde{T}) +\tilde{P} -1 -\frac{\gamma^2}{2}\right],
\end{equation}
\begin{equation}
\label{Eq9} c_2 = 3\gamma\gamma_1,\;\;\;\; c_3 = 2(1-\gamma_1^2).
\end{equation}

The  FS phase has two thermodynamically equivalent phase domains
that can be distinguished by the upper and lower signs ($\pm$) of
some terms in Eqs.~(\ref{Eq6}) and (\ref{Eq7}). The upper sign
describes the domain (labelled bellow again by FS), with $m>0$,
$\mbox{sin}\theta = - 1$, whereas the lower sign describes the
conjunct domain FS$^{\ast}$, where $m < 0$ and $\mbox{sin}\theta =
1$ ; for details, see, Ref.~\cite{Shopova:2005}. Here we consider
one of the two thermodynamically equivalent phase domains, namely,
the domain FS, which is stable for $m >0$ (FS$^{\ast}$ is stable
for $m<0$).  In the absence of external symmetry breaking field,
such ``one-domain approximation" correctly presents the main
thermodynamic properties described by the model (\ref{Eq1}). The
FS phase domain,  given by upper sign Eqs.(\ref{Eq6}) and
(\ref{Eq7}), satisfies the stability conditions: $\gamma M \geq 0$
and
\begin{equation}
\label{Eq10} \kappa (\tilde{T}_s-\tilde{T}) \pm \gamma m - 2
\gamma_1 m^2 \geq 0,
\end{equation}
\begin{equation}
\label{Eq11} 3(1-\gamma_1^2)m^2 +3\gamma\gamma_1m + \tilde{T}^n -1
+\tilde{P} +\kappa\gamma_1(\tilde{T}_s-\tilde{T})
-\frac{\gamma^2}{2} \geq 0.
\end{equation}
These results are valid for $T_f(P) > T_s(P)$, which excludes
any pure superconducting phase ($\mbox{\boldmath$\psi$} \neq 0, m=
0$) in accordance with the available experimental data.

\begin{table*}
\caption{\label{tab:table1}Theoretical results for the location
[$(\tilde{T},\tilde{P})$-{\em reduced} coordinates] of the
tricritical points $A \equiv (\tilde{T}_A, \tilde{P}_A)$ and $B
\equiv (\tilde{T}_B,\tilde{P}_B)$, the critical-end point $C
\equiv (\tilde{T}_C,\tilde{P}_C)$, and the point of temperature
maximum, {\it max} =$(\tilde{T}_m,\tilde{P}_m)$ on the curve
$\tilde{T}_{FS}(\tilde{P})$ of the FM-FS phase transitions of
first and second order (for details, see Sec. III.B). The first
column shows $\tilde{T}_N \equiv \tilde{T}_{(A,B,C,m)}$. The
second column stands for $t_N = t_{(A,B,C,m)}$. The reduced
pressure values $\tilde{P}_{(A,B,C,m)}$ of points A, B, C, and
{\it max} are denoted by $\tilde{P}_N(n)$: $n=1$  stands for  the
linear dependence $T_f(P)$,  $n=2$ - for the nonlinear dependence
of $T_f(P)$ and $t(T)$, corresponding to SFT.}
\begin{ruledtabular}
\begin{tabular}{llll}
 $N$ & $\tilde{T}_N$ & $t_N$& $\tilde{P}_N(n)$ \\ \hline

A & $\tilde{T}_s$ & $ \gamma^2/2$ & $1 - \tilde{T}_s^n+ \gamma^2/2$  \\

B & $\tilde{T}_s + {\gamma^2(2
+\gamma_1)}/{4\kappa(1+\gamma_1)^2}$& $-\gamma^2/4(1+\gamma_1)^2$
&    $1- \tilde{T}_B^n -\gamma^2/4(1+\gamma_1)^2$ \\
C & $\tilde{T}_s + {\gamma^2}/{4\kappa(1+\gamma_1)}$ & $0$ &
$1-\tilde{T}_C^n$   \\
$max$ & $\tilde{T}_s + {\gamma^2}/{4\kappa\gamma_1}$ &
$-\gamma^2/4\gamma_1^2$ & $ 1 - \tilde{T}_m^n -
{\gamma^2}/{4\gamma_1^2}$   \\
\end{tabular}
\end{ruledtabular}
\end{table*}

For $r <0$, and $t > 0$ the model (\ref{Eq1})-(\ref{Eq2}) exhibits
a stable pure superconducting phase ($\phi_1=\phi_2=m=0$,
$\phi_3^2 = -r$)~\cite{Shopova:2005}. This phase may occur in the
temperature domain $T_f(P) < T < T_s$. For systems, where $T_f(0)
\gg T_s$, this is a domain of pressure in a very close vicinity of
$P_0\sim P_c$, where $T_{F}(P)\sim T_f(P)$ decreases to values
lower than $T_s$. Such a situation is described by the model
(\ref{Eq2}) only if $T_s >0$. This case is interesting from the
experimental point of view only when $T_s > 0$ is big enough to
enter in the scope of experimentally measurable temperatures. Up
to date a pure superconducting phase has not been observed within
the accuracy of experiments on the studied metallic compounds. For
this reason, in the reminder of this paper we will  assume that
the critical temperature $T_s$ of the generic superconducting
phase transition is either non-positive $(T_s \leq 0)$, or has a
small positive value which can be neglected in the analysis of the
available experimental data.

The negative values of the critical temperature $T_s$ of the
generic superconducting phase transition are generally possible
and produce a variety of phase diagram topologies (Sec. III.B).
The value of $T_s$ depends on the strength of the interaction,
responsible for the formation of the spin-triplet Cooper pairs,
which means that the P-dependence of $T_s$ specifies the
sensitivity of electron couplings to the crystal lattice
properties. This is an effect which might be included in our
theoretical scheme by introducing some convenient pressure
dependence of $T_s$. To do so we need information either from
experimental data or from a comprehensive microscopic theory.

Usually, $T_s\leq 0$ is interpreted as  a lack of any
superconductivity, but here the same non-positive values of $T_s$
are effectively enhanced to positive values by the interaction
parameter $\gamma$ which triggers the superconductivity up to
superconducting phase transition temperatures $T_{FS}(P)> 0$. This
is  readily seen from Table 1, where we present the reduced
characteristic temperatures on the FM-FS phase transition line
$\tilde{T}_{FS}(\tilde{P})$, calculated from the present theory,
namely, $\tilde{T}_m$ - the maximum of the curve
$\tilde{T}_{FS}(\tilde{P})$ (if available; see Sec. III.B), the
temperatures $\tilde{T}_A$ and $\tilde{T}_B$, corresponding to the
tricritical points $A\equiv (\tilde{T}_A,\tilde{P}_A)$, $B\equiv
(\tilde{T}_B,\tilde{P}_B)$, and the temperature $\tilde{T}_C$
corresponding to the critical-end point $C\equiv
(\tilde{T}_C,\tilde{P}_C)$. The theoretical derivation of the
dependence of the multicritical temperatures $\tilde{T}_A$,
$\tilde{T}_B$, and $\tilde{T}_C$ on $\gamma$, $\gamma_1$,
$\kappa$, and $\tilde{T}_s$, as well as the dependence of
$\tilde{T}_m$ on the same model parameters is outlined in Sec.
III.B. All these temperatures as well as the whole phase
transition line $T_{FS}(P)$ are considerably boosted above $T_s$
owing to positive terms of order $\gamma^2$. If $\tilde{T}_s < 0$,
the superconductivity appears, provided $\tilde{T}_m
> 0$, i.e., when $\gamma^2/4\kappa\gamma_1 >
|\tilde{T}_s|$ (see Table 1).

\subsection{Temperature-pressure phase diagram}

Although the structure of the FS phase is  complicated, some of
the results can be obtained in analytical form. A more detailed
outline of the phase domains, for example, in $T-P$ phase diagram,
can be done by using suitable values of  material parameters in
the free energy~(\ref{Eq2}): $P_0$, $T_{f0}$, $T_s$, $\kappa$,
$\gamma$, and $\gamma_1$. Here we present some of the analytical
results for the phase transition lines and the multi-critical
points. Typical shapes of phase diagrams derived directly from Eq.
(\ref{Eq2}) are given in Figs.1-6. Fig.~1 shows the phase diagram
calculated from Eq.~(\ref{Eq2}) for parameters, corresponding to
the experimental data~\cite{Pfleiderer:2001} for ZrZn$_2$.
Figs.~2-3 show the low-temperature and the high- pressure parts of
the same phase diagram (see Sec. IV for details). Figs.~4-6 show
the phase diagram calculated for the experimental
data~\cite{Saxena:2000, Tateiwa:2001} of UGe$_2$ (see Sec.IV). In
ZrZn$_2$, UGe$_2$, as well as in UCoGe and UIr, critical pressure
$P_c$ exists, where both superconductivity and ferromagnetic
orders vanish.

As in experiments, we find out from our calculation that in the
vicinity of $P_0\sim P_c$, the FM-FS phase transition is of first
order, denoted by the solid line BC in Figs.~2,3,5,6. At lower
pressure the same phase transition is of second order, shown by  the
dotted lines in the same figures. The second order phase
transition line $\tilde{T}_{FS}(P)$ separating the FM and FS
phases is given by the solution of the equation
\begin{equation}
\label{Eq12} \tilde{T}_{FS}(\tilde{P}) = \tilde{T}_s +
\tilde{\gamma_1}t_{FS}(\tilde{P}) +
\tilde{\gamma}[-t_{FS}(\tilde{P})]^{1/2},
\end{equation}
where $t_{FS}(\tilde{P}) = t(T_{FS}, \tilde{P}) \leq 0$,
$\tilde{\gamma} = \gamma/\kappa$, $\tilde{\gamma}_1 =
\gamma_1/\kappa$, and $0 < \tilde{P} < \tilde{P}_B$; $P_B$ is the
pressure corresponding to the multi-critical point B, where the
first-order line $T_{FS}(P)$ terminates, as clearly seen in
Figs.~3 and 6. Eq.~(\ref{Eq12})strictly coincides with the
stability condition for the FM-phase with respect to the
appearance of FS-phase.

Additional information for the shape of the FM-FS phase transition
line can be obtained by the derivative $\tilde{\rho} =
\partial \tilde{T}_{FS}(\tilde{P})/\partial\tilde{P}$, namely,
\begin{equation}
\label{Eq13} \tilde{\rho} = \frac{\tilde{\rho}_s +\tilde{\gamma_1}
- \tilde{\gamma}/2(-t_{FS})^{1/2}}{1 -
n\tilde{T}_{FS}^{n-1}\left[\tilde{\gamma_1} -
\tilde{\gamma}/2[(-t_{FS})^{1/2} \right]},
\end{equation}
where $\tilde{\rho}_s = \partial
\tilde{T}_{s}(\tilde{P})/\partial\tilde{P}$. Note, that
Eq.~(\ref{Eq13}) is obtained from Eqs.~(\ref{Eq5}) and
(\ref{Eq12}).

The shape of the line $\tilde{T}_{FS}(P)$ can vary depending on
the theory parameters (see, e.g., Figs. 2,~5). For certain ratios
of $\tilde{\gamma}$, $\tilde{\gamma}_1$, and values of
$\tilde{\rho}_s$, the curve $\tilde{T}_{FS}(\tilde{P})$ exhibits a
maximum $\tilde{T}_m = \tilde{T}_{FS}(\tilde{P}_m)$, given by the
solution of $\tilde{\rho}(\tilde{\rho}_s, T_m,P_m)=0$. This
maximum is clearly seen in Figs. 5, 6. To locate the maximum we
need to know $\tilde{\rho}_s$. We have already assumed that $T_s$
does not depend on $P$, as explained above,  which from the
physical point of view means that the function $T_s(P)$ is
 flat enough to allow  the approximation $\tilde{T}_{s}\approx
0$ without a substantial error in the results. From our choice of
$P$-dependence of the free energy (\ref{Eq2})parameters, it
follows that $\tilde{\rho}_s = 0 $.

Setting $\tilde{\rho}_s =\tilde{\rho}=
0 $ in Eq.~(\ref{Eq13}), we obtain
\begin{equation}
\label{Eq14} t(T_m,P_m)=
-\frac{\tilde{\gamma}^2}{4\tilde{\gamma}_1^2},
\end{equation}
namely, the value $t_m(T,P) = t(T_m,P_m)$ at the maximum
$T_m(P_m)$ of the curve $T_{FS}(P)$. Substituting $t_m$ back in
Eq.~(\ref{Eq12}) we obtain $T_m$,  and with its help we also obtain
 the pressure $P_m$, both given in Table~1, respectively.

We want to draw the attention to a  particular feature of the
present theory that the coordinates $T_m$ and $P_m$ of the maximum
(point ``{\it max}") on the curve $T_{FS}(P)$, and also the
results of various calculations with the help of Eqs.~(\ref{Eq12})
and (\ref{Eq13}) are expressed in terms of the reduced interaction
parameters $\tilde{\gamma}$ and $\tilde{\gamma}_1$. Thus, using
certain experimental data for $T_m$, $P_m$, as well as
Eqs.~(\ref{Eq12}) and (\ref{Eq13}) for $T_{FS}$, $T_s$ and the
derivative $\rho$ at particular values of the pressure $P$,
$\tilde{\gamma}$ and $\tilde{\gamma}_1$ can be calculated without
any additional information, for example, for the parameter
$\kappa$. This property of the model (\ref{Eq2}) is very useful in
the practical work with the experimental data.
\begin{figure}
\includegraphics[width=7.5cm, height=6.5cm, angle=-90]{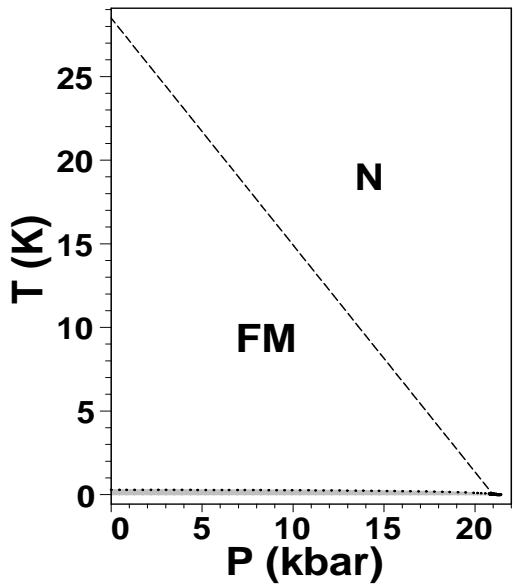}
\caption{\label{fig3} $T-P$ diagram of ZrZn$_2$ calculated for
$T_s=0$, $T_{f0}=28.5$ K, $P_0 = 21$ kbar, $\kappa = 10$,
$\tilde{\gamma} = 2\tilde{\gamma_1} \approx 0.2$, and $n=1$. The
dotted line represents the FM-FS transition and the dashed line
stands for the second order N-FM transition. The dotted line has a
zero slope at $P=0$. The low-temperature and high-pressure domains
of the FS phase are seen more clearly in the following Figs. 2,
3.}
\end{figure}

The conditions for existence of a maximum on the curve $T_{FS}(P)$
can be determined by requiring $\tilde{P}_{m} > 0$ and
$\tilde{T}_m > 0$ and using the respective formulae for these
quantities, shown in Table 1. This {\it max} always occurs in
systems where $T_{FS}(0) \leq 0$ and the low-pressure part of the
curve $T_{FS}(P)$ terminates at $T=0$ for some non-negative
critical pressure $P_{0c}$ (see Sec. III.C). But the maximum  may
occur also for some sets of material parameters, when $T_{FS}(0)>
0$ (see Fig.~2, where $P_m =0$). All these shapes of the line
$T_{FS}(P)$ are described by the model~(\ref{Eq2}). Irrespective
of the particular shape, the curve $T_{FS}(P)$ given by
Eq.~(\ref{Eq12}) always terminates at the tricritical point
(labelled B), with coordinates $(P_B,T_B)$ (see, e.g., Figs. 3,
6).

\begin{figure}
\includegraphics[width=5.5cm, height=7.5cm, angle=-90]{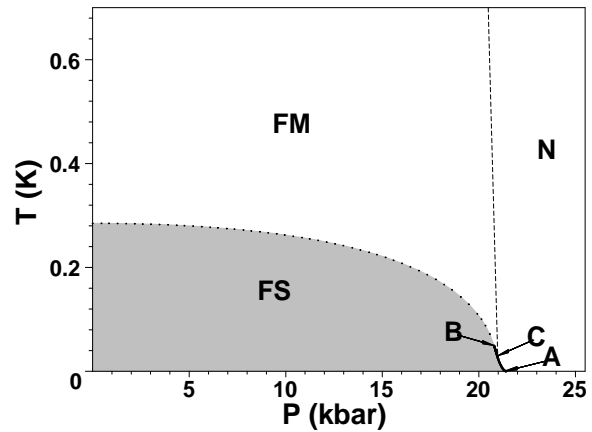}
\caption{\label{fig3} Details of Fig.~1 with expanded temperature
scale. The points A, B, C are located in the high-pressure part
($P\sim P_c\sim 21~\mbox{kbar}$). The $max$ point is at $P\approx
0$~kbar. The FS phase domain is shaded. The dotted line shows the
second order FM-FS phase transition with $P_m \approx 0$. The
solid straight line BC shows the fist-order FM-FS transition for
$P > P_B$. The quite flat solid line AC shows the first order N-FS
transition (the lines BC and AC are more clearly seen in Fig.~3.
The dashed line stands for the second order N-FM transition.}
\end{figure}

At pressure $P > P_B$ the FM-FS phase transition is of first order
up to the critical-end point C. For $P_B < P < P_C$ the FM-FS
phase transition is given by the straight line BC; see,
Figs.~3, 6. All three phase transition lines - N-FM, N-FS,
and FM-FS -  terminate at point C. For $P > P_C$ the FM-FS phase
transition occurs on a rather flat, smooth line of equilibrium
transition of first order up to a second tricritical point A with
$P_A \sim P_0$ and $T_A \sim 0$. Finally, the third transition
line terminating at the point C describes the second order phase
transition N-FM. The reduced temperatures $\tilde{T}_N$ and
pressures $\tilde{P}_N$, $N$ = (A, B, C, {\it max}) at the three
multi-critical points (A, B, C) and the maximum $T_m(P_m)$ are
given in Table 1. Note that, for any set of material parameters,
$T_A < T_C < T_B < T_m$ and $P_m<P_B<P_C<P_A$.

There are other types of phase diagrams, resulting from  the model
 (\ref{Eq2}). For negative values
of the generic superconducting temperature $T_s$ several other
topologies of the $T-P$ diagram can be outlined. The results for
the multi-critical points, presented in Table 1 show that when
$T_s$ lowers below $T=0$, also $T_C$ decrease, firstly to zero,
and then to negative values. When $T_C=0$ the direct N-FS phase
transition of first order disappears and the point C becomes a
very special zero-temperature multi-critical point. As seen from
Table 1, this happens for $T_s = -
\gamma^2T_f(0)/4\kappa(1+\gamma_1)$. The further decrease of $T_s$
causes the point C to fall below the zero temperature and then the
zero-temperature phase transition of first order near $P_c$ splits
into two zero-temperature phase transitions: a second order N-FM
transition and a first order FM-FS transition, provided $T_B$
still remains positive.

At lower $T_s$ also the
point $B$ falls below $T=0$ and the FM-FS phase transition becomes
entirely of second order. For very extreme negative values of $T_s$,
 a very large pressure interval below $P_c$ may occur where the
FM phase is stable up to $T=0$. Then the line $T_{FS}(P)$ will
exist only for relatively small pressure values $(P \ll P_c)$.
This shape of the stability domain of the FS phase is also
possible in real systems.

\begin{figure}
\includegraphics[width=6.5cm, height=7.5cm, angle=-90]{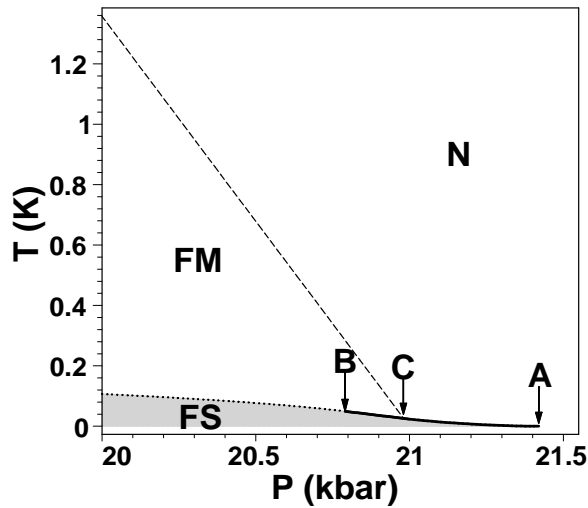}
\caption{\label{fig3} High-pressure part of the phase diagram of
ZrZn$_2$, shown in Fig.~1. The thick solid lines AC and BC show
the first-order transitions N-FS, and FM-FS, respectively. Other
notations are explained in Figs. 1, 2.}
\end{figure}

\subsection{Quantum phase transitions}

The effects of quantum correlations on the phase transition
properties at ultra-low and zero temperature, known as quantum
phase transitions~\cite{Shopova:2003}, can  also be described by
the free energy~(\ref{Eq1}), as mentioned above. The
time-dependent quantum fluctuations, which describe the intrinsic
quantum dynamics of spin-triplet ferromagnetic superconductors,
are not included in the model~(\ref{Eq1}),  but some basic
properties of the quantum phase transitions can be outlined within
the classical limit, namely, by the free energy model (\ref{Eq1})
-- (\ref{Eq2}).

Generally, both classical (thermal) and quantum fluctuations are
investigated by the method of the renormalization group
(RG)~\cite{Uzunov:1993}, which is specially intended to treat the
generalized action of system, where the order parameter fields
fluctuate in time $t$ and space $\vec{x}$~\cite{Uzunov:1993,
Shopova:2003}. These effects, which are beyond the scope of the
paper, lead either to precise treatment of the narrow critical
region in a very close vicinity of second order phase transition
lines, or to  a fluctuation-driven change of the phase transition
order. But, the thermal fluctuation and quantum correlation
effects on the thermodynamics of a given system can be
unambiguously estimated only after the results from the
counterpart simpler theory, where these effects are not present,
are known and, hence, the distinction in the thermodynamic
properties predicted by the respective variants of the theory can
be established. Here we show that the basic low and ultra-low
temperature properties of the spin-triplet FSs, as given by the
preceding experiments, are derived from the model (1) without any
account of fluctuation phenomena and quantum correlations. The
latter might be of use in a more detailed consideration of the
close vicinity of quantum critical points in the phase diagrams of
spin-triplet FSs. Here we show that the theory predicts quantum
critical phenomena only for quite particular physical conditions
whereas the low- and zero-temperature phase transitions of the
first order are favored by both symmetry arguments and detailed
thermodynamic analysis.

There is a number of experimental~\cite{Huy:2007, Uhlarz:2004} and
theoretical~\cite{Nevidomskyy:2005, Uzunov:2006, Belitz:2005}
studies on the problem of quantum phase transitions in
unconventional ferromagnetic superconductors, including the
mentioned intermetallic compounds. Some of them are based on
different theoretical schemes, and do not refer to the model
(\ref{Eq1}). Other, for example, those in Ref.~\cite{Uzunov:2006}
reported results about the thermal and quantum fluctuations,
described by the model (\ref{Eq2}) before the comprehensive
knowledge of the results from the basic treatment, given by the
present investigation. In such cases one could not be sure about
the correct interpretation of the results from the RG and the
possibilities for their application to particular zero temperature
phase transitions. Here we present basic results for the
zero-temperature phase transitions, described by the model
(\ref{Eq2}).

\begin{figure}
\includegraphics[width=6.5cm, height=7.5cm, angle=-90]{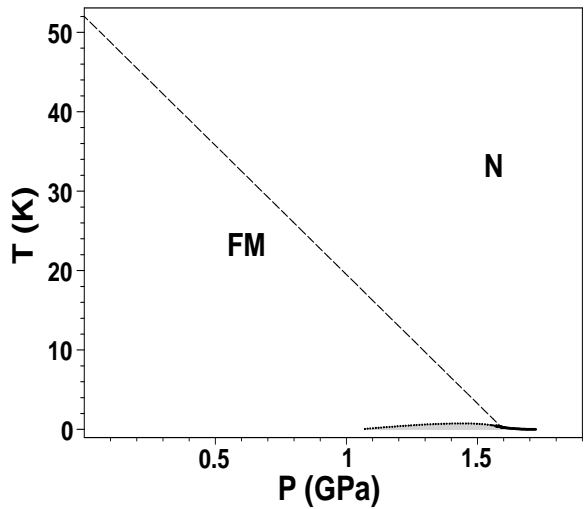}
\caption{\label{fig3} $T-P$ diagram of UGe$_2$ calculated taking
$T_s=0$, $T_{f0}=52$ K, $P_0 = 1.6$ GPa, $\kappa = 4$,
$\tilde{\gamma} = 0.0984$, $\tilde{\gamma_1} = 0.1678$, and $n=1$.
The dotted line represents the FM-FS transition and the dashed
line stands for the N-FM transition. The low-temperature and
high-pressure domains of the FS phase are seen more clearly in the
following Figs.~5,~6.}
\end{figure}

The RG investigation~\cite{Uzunov:2006} has demonstrated up to two
loop order of the theory that the thermal fluctuations of the
order parameter fields re-scale the model (\ref{Eq2}) in way which
corresponds to first order phase transitions in magnetically
anisotropic systems. This result is important for the metallic
compounds, we consider here, because in all of them  magnetic
anisotropy is present. The uniaxial magnetic anisotropy in
ZrZn$_2$ is much more weak than in UGe$_2$ but cannot be neglected
when fluctuation effects are accounted for. Owing to the
particular symmetry of the model (\ref{Eq2}), for the case of
magnetic isotropy (Heisenberg symmetry), the  RG study reveals an
entirely new class of (classical) critical behavior. Besides, the
different spatial dimensions of the superconducting and magnetic
quantum fluctuations imply a lack of stable quantum critical
behavior even when the system is completely magnetically
isotropic. The pointed arguments and preceding results lead to the
reliable conclusion that the phase transitions, which have already
been proven to be first order in the lowest order approximation,
where thermal and quantum fluctuations are neglected, will not
undergo a fluctuation-driven change of the phase transition order
from first to second. Such picture is described below,  in Sec.
IV, and it corresponds to the behavior of real compounds.

Our results definitely show that the quantum phase
transition near $P_c$ is of first order. This is valid for the
whole N-FS phase transition below the critical-end point C, as
well as the straight line BC. The simultaneous effect of thermal
and quantum fluctuations do not change the order of the N-FS
transition, and it is  unlikely to suppose that thermal
fluctuations of the superconductivity field
$\mbox{\boldmath$\psi$}$ can ensure a fluctuation-driven change of
the order of the FM-FS transition along the line BC. Usually, the
fluctuations of $\mbox{\boldmath$\psi$}$ in low temperature
superconductors are small and slightly influence the phase
transition in  the very narrow critical region in the vicinity of
phase transition point. This effect is very weak and can hardly
be observed in any experiment on low-temperature superconductors.
Besides, the fluctuations of the magnetic induction
$\mbox{\boldmath$B$}$ always tend to a fluctuation-induced first
order phase transition rather than to the opposite effect - the
generation of magnetic fluctuations with infinite correlation
length at the equilibrium phase transition point and, hence, a
second order phase transition ~\cite{Uzunov:1993, Halperin:1974}.
Thus we can  reliably conclude that the first order phase
transitions at low-temperatures, represented by the lines BC and
AC in vicinity of $P_c$ do not change their order as a result of
thermal and quantum  fluctuations.

\begin{figure}
\includegraphics[width=5.5cm, height=7.5cm, angle=-90]{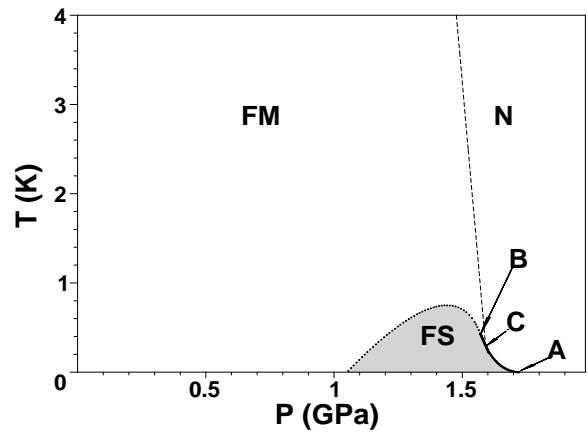}
\caption{\label{fig3} Low-temperature part of the $T-P$ phase
diagram of UGe$_2$, shown in Fig.~4. The points A, B, C are
located in the high-pressure part ($P\sim P_c\sim 1.6$ GPa). The
FS phase domain is shaded. The thick solid lines AC and BC show
the first-order transitions N-FS, and FM-FS, respectively. Other
notations are explained in Figs. 1, 2.}
\end{figure}

Quantum critical behavior  for continuous phase transitions in
spin-triplet ferromagnetic superconductors with magnetic
anisotropy can therefore be observed at other zero-temperature
transitions, which may occur in these systems far from the
critical pressure $P_c$. This is possible when $T_{FS}(0) = 0$ and
the $T_{FS}(P)$ curve terminates at $T=0$ at one or two quantum
(zero-temperature) critical points: $P_{0c} < P_m$ - ``lower
critical pressure", and $P_{0c}^{\prime}>P_m$ -- ``upper critical
pressure." To obtain these critical pressures one should solve
Eq.~(\ref{Eq12}) with respect to $P$, provided $T_{FS}(P) = 0$,
$T_m > 0$ and $P_m>0$, namely, when the continuous function
$T_{FS}(P)$ exhibits a maximum. The critical pressure
$P_{0c}^{\prime}$ is bounded in the relatively narrow interval
($P_m,P_B$) and can appear for some special sets of material
parameters ($r,t,\gamma,\gamma_1$). In particular, as our
calculations show, $P_{0c}^{\prime}$ do not exist for $T_s \geq
0$.

The analytical calculation of the critical pressures $P_{0c}$ and
$P_{0c}^{\prime}$ for the general case of $T_s \neq 0$ leads to
very complex conditions for the appearance of  second critical
field $P_{0c}^{\prime}$. The correct treatment for $T_s
\neq 0$ can be performed within the  two-domain picture of
the phase FS. The complete  study of this case is beyond our
aims, but here we will  illustrate our arguments by
investigation of the conditions, under which the critical pressure
$P_{oc}$ occurs in systems with $T_s \approx 0$. Moreover, we will
present the general result for $P_{0c} \geq 0$ and
$P_{0c}^{\prime} \geq 0$ in systems where $T_s \neq 0$.

\begin{figure}
\includegraphics[width=5.5cm, height=7.5cm, angle=-90]{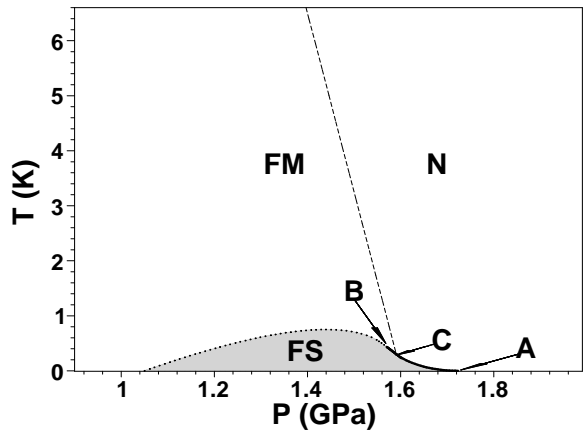}
\caption{\label{fig3} High-pressure part of the phase diagram of
UGe$_2$, shown in Fig. 4. Notations are explained in Figs.~1, 2,
4, 5.}
\end{figure}

Setting $T_{FS}(P_{0c}) = 0 $ in Eq.~(\ref{Eq12}) we obtain the
following quadratic equation
\begin{equation}
\label{Eq15}  \tilde{\gamma}_1 m^{2}_{0c} - \tilde{\gamma}m_{0c} -
\tilde{T}_s = 0
\end{equation}
for the reduced magnetization
\begin{equation}
\label{Eq16} m_{0c} =  [-t(0,\tilde{P}_{oc})]^{1/2} =
(1-\tilde{P}_{0c})^{1/2}
\end{equation}
and, hence, for $\tilde{P}_{0c}$. For $T_s \neq 0$ the
Eqs.~(\ref{Eq15}) and (\ref{Eq16}) have two solutions with respect
to $\tilde{P}_{0c}$. For some sets of material parameters these
solutions satisfy the physical requirements for $P_{0c}$ and
$P_{0c}^{\prime}$ and can be identified with the critical
pressures. The conditions for existence of $P_{0c}$ and
$P_{0c}^{\prime}$ can be obtained either by analytical
calculations, or by numerical analysis for
particular values of the material parameters.

For $T_s=0$, the trivial solution $\tilde{P}_{0c} = 1$ corresponds
to $P_{0c} = P_0
> P_B$ and, hence, does not satisfy the physical requirements. The
second solution
\begin{equation}
\label{Eq17} \tilde{P}_{0c} =  1 -
\frac{\tilde{\gamma}^2}{\tilde{\gamma}^2_1}
\end{equation}
is positive for
\begin{equation}
\label{Eq18} \frac{\gamma_1}{\gamma} \geq 1
\end{equation}
and, as shown below,  it gives the
location of the quantum critical point $(T=0, P_{0c} < P_m)$. At
this quantum critical point, the equilibrium magnetization
$m_{0c}$ is described by $m_{0c} = \gamma/\gamma_1$ and is twice
bigger that the magnetization $m_{m} =
\gamma/2\gamma_1$~\cite{Shopova:2005} at the maximum of the curve
$T_{FS}(P)$.

 By taking $\tilde{P}_m$ from Table 1, we can show that the solution
(\ref{Eq17}) satisfies the condition $P_{0c} < \tilde{P}_m$ for $n=1$, if

\begin{equation}
\label{Eq19} \gamma_1 < 3\kappa
\end{equation}

and for $n=2$ (SFT case), when

\begin{equation}
\label{Eq20} \gamma < 2\sqrt{3}\kappa.
\end{equation}
Finally, we determine the conditions under which the maximum $T_m$
of the curve $T_{FS}(P)$ occurs at non-negative pressures. For
$n=1$, we obtain that $P_m \geq 0$ for $n=1$, if

\begin{equation}
\label{Eq21} \frac{\gamma_1}{\gamma} \geq \frac{1}{2}\left(1 +
\frac{\gamma_1}{\kappa}\right)^{1/2},
\end{equation}
whereas for $n=2$,  the condition is

\begin{equation}
\label{Eq22} \frac{\gamma_1}{\gamma} \geq
\frac{1}{2}\left(1+\frac{\gamma^2}{4\kappa^2}\right)^{1/2}.
\end{equation}

Obviously, the conditions~(\ref{Eq18})-(\ref{Eq22}) are compatible with
one another. The condition (\ref{Eq21}) is weaker than the
condition~(\ref{Eq18}), provided the inequality~(\ref{Eq19})
is satisfied. The same is valid for the condition (\ref{Eq22}),  if
the inequality (\ref{Eq20}) is valid. In Sec. IV we will show that
these theoretical predictions are confirmed by the experimental
data.

Doing in the same way the analysis of Eq.~(\ref{Eq12}) some
results may easily be obtained  for $T_s \neq 0$. In this more
general case the Eq.~(\ref{Eq12}) has two nontrivial solutions
that yield two possible values of the critical pressure.

\begin{equation}
\label{Eq23} \tilde{P}_{0c(\pm)} = 1 -
\frac{\gamma^2}{4\gamma_1^2}\left[1 \pm
\left(1+\frac{4\tilde{T}_s\kappa\gamma_1}{\gamma^2}\right)^{1/2}\right]^2.
\end{equation}
The relation $\tilde{P}_{0c(-)} \geq \tilde{P}_{0c(+)}$ is always
true. Therefore, to have both $\tilde{P}_{0c(\pm)} \geq 0$, it is
enough to require that $\tilde{P}_{0c(+)} \geq 0$. Having in mind
that for the phase diagram shape, we study, $\tilde{T}_m > 0$, and
according to the result for $\tilde{T}_m$ in Table 1,  this leads
to inequality  $\tilde{T}_s> -\gamma^2/4\kappa\gamma_1$. So, we
obtain that $\tilde{P}_{0c(+)} \geq 0$ will exist, if

\begin{equation}
\label{Eq24} \frac{\gamma_1}{\gamma} \geq 1 +
\frac{\kappa\tilde{T}_s}{\gamma},
\end{equation}
which generalizes the condition (\ref{Eq18}).

Now we can identify the pressure $P_{0c(+)}$ with the lower
critical pressure $P_{0c}$, and $P_{0c(-)}$ with the upper
critical pressure $P_{0c}^{\prime}$. Therefore, for wide
variations of the material parameters, the theory (\ref{Eq1})
describes a quantum critical point $P_{oc}$ that exists, provided
the condition (\ref{Eq24}) is satisfied. The quantum critical
point $(T=0,P_{0c})$  exists in UGe$_2$ and, perhaps, in other
$p$-wave ferromagnetic superconductors, for example, in UIr.

Our results predict the appearance of second critical pressure -
the upper critical pressure $P_{oc}^{\prime}$ that  exists under
more restricted conditions and, hence, can be observed for those
systems, where $T_s < 0$. As mentioned in Sec. III.B, for very
extreme negative values of $T_s$, when $T_B < 0$, the upper
critical pressure $P_{0c}^{\prime}>0$ occurs, whereas the lower
critical pressure $P_{0c}>0$  does not appear. But especially this
situation should be investigated in a different way, namely, one
should keep $T_{FS}(0)$ different from zero in Eq.~(\ref{Eq12}),
and consider a form of the FS phase domain in which the curve
$T_{FS}(P)$ terminates at $T=0$ for $P_{0c}^{\prime}>0$,
irrespective of whether the maximum $T_m$ exists or not. In such
geometry of the FS phase domain, the maximum $T_(P_m)$ may exist
only in quite unusual cases, if it exists at all.

The quantum and thermal fluctuation phenomena in the vicinities of
the two critical pressures $P_{0c}$ and $P_{0c}^{\prime}$ need a
non-standard RG treatment because they are related with the
fluctuation behavior of the superconducting field
$\mbox{\boldmath$\psi$}$ far below the ferromagnetic phase
transitions, where the magnetization $\mbox{\boldmath$M$}$ does
not undergo significant fluctuations and can be considered
uniform. The presence of uniform magnetization produces couplings
of $\mbox{\boldmath$M$}$ and $\mbox{\boldmath$\psi$}$ which are
not present in previous RG studies and need a special
analysis.

\section{Application to metallic compounds}

\subsection{Theoretical outline of the phase diagram}

In order to apply the above displayed theoretic calculations,
following from the free energy~(\ref{Eq2}),  for the outline of
$T-P$ diagram of any material, we need information about the
values of  $P_0$, $T_{f0}$, $T_s$, $\kappa$ $\gamma$, and
$\gamma_1$. The temperature $T_{f0}$ can be obtained directly from
the experimental phase diagrams. The pressure $P_0$ is either
identical  or very close to the critical pressure $P_c$, for which
the N-FM phase transition line terminates at $T \sim 0$. The
temperature $T_s$ of the generic superconducting transition is not
available from the experiments because, as mentioned above pure
superconducting phase not coexisting with ferromagnetism has been
not observed.  This can be considered as an indication that  $T_s$
is small and does not produce a measurable effect. So the generic
superconducting temperature will  be estimated on the basis of
following general arguments. For $T_f(P)> T_s$ we must have
$T_s(P) =0$ at $P \geq P_c$, where $T_f(P) \leq 0$ and for $0 \leq
P\leq P_0$, $T_s < T_C$. Therefore for materials where  $T_C$ is
too small to be observed experimentally, $T_s$ can be ignored.

As far as  the shape of FM-FS transition line is well described by
Eq.~(\ref{Eq12}) we will make use of additional data from
available experimental phase diagram for ferromagnetic
superconductors. For example, in ZrZn$_2$ these are the observed
values of $T_{FS}(0)$ and the slope $\rho_0 \equiv [\partial
T_{FS}(P)/\partial P]_0 =(T_{f0}/P_0)\tilde{\rho}_0 $ at $P=0$.
For UGe$_2$, where a maximum is observed on the phase transition
line, we can use the experimental values for  $T_m$, $P_m$, and
also for $P_{0c}$. The interaction parameters $\tilde{\gamma}$ and
$\tilde{\gamma_1}$ are derived  using Eq. (\ref{Eq12}) and the
expressions for $\tilde{T}_m$, $\tilde{P}_m$, and
$\tilde{\rho}_0$, see Table~1. The parameter  $\kappa$ is chosen
by fitting the expression for the critical-end point $T_C$.

\subsection{Diagram of ZrZn$_2$}

Experiments  for ZrZn$_2$~\cite{Pfleiderer:2001} give the
following values:  $T_{f0} = 28.5$ K, $T_{FS}(0) = 0.29$ K, $P_0
\sim P_c = 21$ kbar. The  curve $T_F(P)\sim T_f(P)$ is almost a
straight line, which directly indicates that  $n=1$ is adequate in
this case for the description of the $P$-dependence. The slope for
$T_{FS}(P)$ at $P=0$ is estimated from the condition that its
magnitude should not exceed $T_{f0}/P_c \approx 0.014$ as we have
assumed that the transition line is straight one, so in result we
have $-0.014 <\rho \leq 0$. This ignores the presence of a
maximum.  The available experimental data for  ZrZn$_2$, do not
give clear indications whether a maximum at ($T_m,P_m$) exists. If
there is a maximum  at $P=0$ this will mean  $\rho_0 = 0$, and a
maximum with $T_m \sim T_{FS}(0)$ and $P_m \ll P_0$ will give
$0\leq \rho_0 < 0.005$. We may choose that  $\rho_0 = 0$, then
$\tilde{\gamma} \approx 0.02$ and $\tilde{\gamma}_1  \approx
0.01$,  similar values for the parameters are obtained  for any
$|\rho_0| \le 0.003$.

The multi-critical points A and C cannot be distinguished
experimentally as the  experimental accuracy
\cite{Pfleiderer:2001} is less than $\sim 25$ mK in the
high-pressure domain ($P\sim 20-21$ kbar).   Having this mind we
can propose as initial assumption that $T_C \sim 10$ mK, which
corresponds to $\kappa \sim 10$, see Table~1. We have calculated
the $T-P$ diagram using $\rho_0 = 0$ and $\rho_0 = 0.003$, as well
as  the other parameters,  discussed above. The differences,
obtained in these two cases, are negligible.

In Fig. 1 we show the phase diagram of ZrZn$_2$ calculated
directly from the free energy (\ref{Eq2}) for $n=1$, the above
mentioned values of $T_s$, $P_0$, $T_{f0}$, $\kappa$, and values
of $\tilde{\gamma} \approx 0.2$ and $\tilde{\gamma}_1 \approx 0.1$
which ensure $\rho_0 \approx 0$. The obtained   coordinates of
characteristic points are the following: $P_A\sim P_c= 21.42$
kbar,  $T_A=T_F(P_c)=T_{FS}(P_c) = 0$ K; $P_B =20.79$ kbar, $T_B
=0.0495$ K, and $P_C = 20.98$ kbar, $T_C =0.0259$ K .

In Fig. 2, the  low-$T$ region is seen in more detail  with  A, B,
C points, respectively. The dotted $T_{FS}(P)$ curve, represents
the phase transition line FM-FS  of second order and  has a
maximum $T_m=0.290$ K at $P = 0.18$ kbar, which is slightly above
$T_{FS}(0) = 0.285$~K. The straight solid line BC in Fig.~2 shows
the first order FM-FS phase transition which occurs for $P_B < P <
P_C$. The solid AC line shows the first order N-FS phase
transition and the dashed line stands for the N-FM phase
transition of second order.

Although the expanded temperature scale in Fig.~2, the difference
$[T_m-T_{FS}(0)] = 5$ mK can hardly be seen. The calculations,
both analytical and numerical, lead to the conclusion that if we
want to  locate the point {\it max} exactly at $P=0$  we have to
take  values of $\tilde{\gamma}$ and $\tilde{\gamma}_1$ with
accuracy up to $10^{-4}$. This may be considered as an indication
that  the {\it max} for parameters corresponding to ZrZn$_2$ is
very sensitive to very small changes of parameters
$\tilde{\gamma}$ and $\tilde{\gamma}_1$ around the values $0.2$
and $0.1$, respectively. Our initial idea was to present a diagram
with $T_m=T_{FS}(0) = 0.29$~K and $\rho_0 = 0$, with the point of
maximum being exactly  at $P=0$, but the final phase diagram
slightly departs from this picture because of the mentioned
sensitivity of the result on the values of the interaction
parameters $\gamma$ and $\gamma_1$.

We have also calculated the  theoretical phase diagram of ZrZn$_2$
for $\rho_0 = 0.003$ and initial values of $\tilde{\gamma}$ and
$\tilde{\gamma}_1$ which differ from $\tilde{\gamma} =
2\tilde{\gamma}_1 = 0.2$ only by numbers of order
$10^{-3}-10^{-4}$~\cite{Cottam:2008}. The obtained values for the
coordinates  of the maximum are: $T_m = 0.301$ K at
$P_m=6.915$~kbar.  This result confirms the mentioned sensitivity
of the location of the maximum $T_m$ towards slight variations of
the material parameters. Experimental investigations of this low
temperature-low pressure region with higher accuracy may help to
locate this maximum with better precision.

We show by Fig.~3 the high-pressure part of  same phase diagram .
As far as the experimental  phase diagram~\cite{Pfleiderer:2001}
has  restricted accuracy in this range of temperatures and
pressures, it is important to draw the attention to the first
order phase transitions, shown by solid lines BC (FM-FS) and AC
(N-FS) as this interesting topology of the phase diagram of
ZrZn$_2$ in the high-pressure domain ($P_B < P < P_A$) is not seen
in the experiments. In fact the line AC is quite flat but not
straight as the line BC. We hope that the detailed theoretical
results for those parts of the phase diagram, where experimental
data are lacking,  may stimulate the further experiments in those
ranges of pressure and temperature. The theoretical diagram, shown
in Fig.~1, looks almost the same as the experimental one as far as
there the details  are smeared. Note, that the
theoretical diagram is a direct result of  calculations with the
free energy  (\ref{Eq2}) and the proposed P-dependence of its
parameters,  without any approximations and simplifying
assumptions.

The theoretical phase diagram reflects  well  the main features of
the experimental behavior \cite{Pfleiderer:2001}, including the
claimed change in the order of the FM-FS phase transition at
relatively high $P$. Within the present model the N-FM transition
is of second order up to $P_C \sim P_c$. Experiments give that the
first order N-FM transition continues to much lower $P$ values,
and if this result is reliable,  the theory can be easily modified
to include this by a change of sign of $b_f$. Then a  new
tricritical point will appear, located at some $P_{tr} < P_C$ on
the N-FM transition line. Since $T_C>0$, a direct N-FS phase
transition of first order is predicted in accord with conclusions
from de Haas--van Alphen experiments \cite{Kimura:2004} and some
theoretical studies \cite{Uhlarz:2004}. Such transition may not
occur in other cases where $T_C=0$. In spin-fluctuation
theory($n=2$) the diagram topology remains the same but points B
and C are slightly shifted to higher $P$, typically by about $0.01
- 0.001$ kbar.

\subsection{Diagram of UGe$_2$}

The experimental data  for UGe$_2$ give $T_{f0} = 52$ K, $P_c=1.6$
GPa ($\equiv 16$ kbar), $T_m = 0.75$ K, $P_m\approx 1.15$ GPa,
\cite{Saxena:2000},  and $P_{0c} \approx 1.05$
GPa~\cite{Saxena:2000, Huxley:2001,Tateiwa:2001, Harada:2007}. We
use again $n=1$ in the pressure dependence of $T_f(P)$
(\ref{Eq4}). From the above cited values for $T_m$ and $P_{0c}$ we
obtain $\tilde{\gamma} \approx 0.0984$ and $\tilde{\gamma_1}
\approx 0.1678$. The temperature $T_C \sim 0.1$ K corresponds to
$\kappa \sim 4$.

In Fig.~4 we show the calculated with these initial parameters
$T-P$ diagram of UGe$_2$. We have assumed that $T_s=0$ in
compliance with the arguments in Sec.~III.A. The location of
points A, B, C on the phase diagram is given by the following
values: $T_A=0$ K, $P_A = 1.723$ GPa, $T_B=0.481$ K, $P_B = 1.563$
GPa, $T_C=0.301$ K, and $P_C=1.591$ GPa. Figs.~5 and 6 show the
details of low-temperature and the high-pressure parts of this
phase diagram, respectively.

Generally the main features of experimental curves are properly
modelled by the theoretical T-P diagram of UGe$_2$, although there
is some discrepancy in the numerical values for $P_m$
corresponding to the maximum. Theoretically it is calculated as
$\sim 1.44$ GPa in Fig.~4 which is about 0.3 GPa higher than found
experimentally~\cite{Tateiwa:2001, Harada:2007}. If the
experimental measurement are precise and give correct results for
$P_m$, this difference may be attributed to the so-called ($T_x$)
metamagnetic phase transition in UGe$_2$. It manifests in  an
abrupt change of the magnetization value in the vicinity of $P_m$.
The  possibility of metamagnetic transition is not considered in
our model, but we may suppose that it will change the shape of
$T_{FS}(P)$ making  more symmetric the slope of curve  around
$P_m$ and lowering it together with $P_B$ and $P_C$. We have tried
to obtain a lower $P_m$ value, keeping  $T_m$ unchanged within the
model (\ref{Eq2}), but this leads  to a change of $P_{c0}$ to a
value that disagrees with experiment. In spin-fluctuation approach
$(n=2)$ the multi-critical points are located at slightly higher
$P$ (by about 0.01 GPa), as for ZrZn$_2$. Therefore, the results
from spin-fluctuation theory are slightly worse than the results
obtained from the usual linear approximation ($n=1$) for the
parameter $t$.

In principle, the location of $P_m$ both for ZrZn$_2$ and UGe$_2$
may be tuned better within the present model~(\ref{Eq1}), if we
take into account the anisotropy effects. Previous
calculations~\cite{Shopova:2005} show that the inclusion,
especially of spin-triplet Cooper pair anisotropy (the $u_s$-term
in the Eq.~(\ref{Eq1})), changes the values of coordinates of
points A, B, C and \emph{max}. The problem is that the anisotropy
in the present status of knowledge may serve only as a fitting
parameter, because no proper information from microscopic theories
is available both for its sign and magnitude. But such fitting
from general considerations can obscure other effects, as the
mentioned above metamagnetic transition in UGe$_2$, and even lead
to some wrong interpretations when comparing the experimental and
theoretical curves.

\subsection{Two types of ferromagnetic superconductors with
spin-triplet electron pairing}

The estimates for UGe$_2$ give that  $\gamma_1\kappa\approx 1.9$,
so the condition for $T_{FS}(P)$ to have a maximum found from Eq.
(\ref{Eq12}) is satisfied. As we have discussed for ZrZn$_2$, the
location of this maximum can be hard to fix accurately in
experiments. However, $P_{c0}$ can be more easily distinguished,
as in the UGe$_2$ case. Then we have a well-established quantum
(zero-temperature) phase transition of second order, i.e., a
quantum critical point at some critical pressure $P_{0c} \geq 0$.
As shown in Sec.~III.C, under special conditions the quantum
critical points could be two: at the lower critical pressure
$P_{0c} < P_m$ and the upper critical pressure $P_{0c}^{\prime} <
P_m$. This type of behavior in systems with $T_s=0$ (as UGe$_2$)
occurs when the criterion (\ref{Eq18}) is satisfied. Such systems
(which we label as U-type) are essentially different from those
such as ZrZn$_2$ where $\gamma_1 < \gamma$ and hence $T_{FS}(0) >
0$. In this latter case (Zr-type compounds), a maximum $T_m> 0$
may sometimes occur, as discussed earlier. We note that the ratio
$\gamma/\gamma_1$ reflects a balance effect of two interactions
between $\mbox{\boldmath$\psi$}$-$\mbox{\boldmath$M$}$. When the
trigger interaction, represented by $\gamma$, prevails, the
Zr-type behavior is found where superconductivity exists at $P=0$.
The same ratio can be expressed as $\gamma_0/\delta M_0$, which
emphasizes that the ground state value of the magnetization at
$P=0$ is also relevant. Alternatively, one may refer to these two
basic types of spin-triplet FSs as ``type I'' (for example, for
the ``Zr-type compounds''), and ``type II'' - for the U-type FS
compounds.

As we see from this classification, the two types of spin-triplet
ferromagnetic superconductors have quite different phase diagram
topologies although some fragments have common features. The same
classification can include systems with $T_s\neq 0$ but in this
case one should use the more general criterion (\ref{Eq24}).

\subsection{Other compounds}

In URhGe, $T_{f}(0) \sim 9.5$~K and $T_{FS}(0) = 0.25$ K and,
therefore, as in ZrZn$_2$,here the spin-triplet superconductivity
appears at ambient pressure deeply in the ferromagnetic phase
domain~\cite{Aoki:2001, Hardy1:2005, Hardy2:2005}. Although some
similar structural and magnetic features  with UGe$_2$ the results
in Ref.~\cite{Hardy2:2005} of measurements under high pressure
show that, unlike the behavior of ZrZn$_2$ and UGe$_2$, the
ferromagnetic phase transition temperature $T_{F}(P)\sim T_{f}(P)$
has a slow linear increase up to $140$~kbar without any
experimental indications that the N-FM transition line may change
its behavior at higher pressures and show a negative slope in
direction of low temperature up to a quantum critical point
$T_{F}=0$ at some critical pressure $P_c$. Such  behavior of the
generic ferromagnetic phase transition temperature cannot be
explained by our initial assumption for the function $T_f(P)$
which was intended to explain phase diagrams where the
ferromagnetic order is depressed by the pressure and vanishes at
$T=0$ for some critical pressure $P_c$. The $T_{FS}(P)$ line of
URhGe shows a clear monotonic negative slope to $T=0$ at pressures
above $15$ kbar and the extrapolation~\cite{Hardy2:2005} of the
experimental curve $T_{FS}(P)$ tends a quantum critical point
$T_{FS}(P_{oc}^{\prime})=0$ at $P_{0c} \sim 25-30$ kbar. Within
the framework of the phenomenological theory (\ref{Eq2}), this
$T-P$ phase diagram can be explained after a modification of
$T_f(P)$-dependence is made, and by introducing a convenient
nontrivial pressure dependence of the interaction parameter
$\gamma$. Such modifications of the present theory are possible
and follow from important physical requirements related with the
behavior of the $f$-band electrons in URhGe. Unlike UGe$_2$, where
the pressure increases the hybridization of the $5f$ electrons
with band states leading to a suppression of the spontaneous
magnetic moment $M$, in URhGe this effects is followed by a
stronger effect of enhancement of the exchange coupling due to the
same hybridization, and this effect leads to the slow but stable
linear increase of the function $T_F(P)$ \cite{Hardy2:2005}. These
effects should be taken into account in the modelling of the
pressure dependence of the parameters of the theory
(\ref{Eq2})when applied to URhGe.

Another ambient pressure FS phase has been observed in experiments
on UCoGe~\cite{Huy:2007}. Here the experimentally derived slopes
of the functions $T_{F}(P)$ and $T_{FS}(P)$ at relatively small
pressures are opposite compared to those for URhGe and, hence, the
$T-P$ phase diagram of this compound can be treated within the
present theoretical scheme without substantial modifications.

Like in UGe$_2$, the FS phase in UIr~\cite{Kobayashi:2006} is
embedded in the high-pressure/low-temperature part of the
ferromagnetic phase domain near the critical pressure $P_c$ which
means that UIr is certainly a U-type compound (see Sec. IV.D). In
UGe$_2$ there is one meta-magnetic phase transition between two
ferromagnetic phases (FM1 and FM2), in UIr there are three
ferromagnetic phases and the FS phase is located in the
low-$T$/high-$P$ domain of the third of them - the phase FM3.
There are two meta-magnetic-like phase transitions: FM1-FM2
transition which is followed by a drastic decrease of the
spontaneous magnetization when the lower-pressure phase FM1
transforms to FM2, and a peak of the ac susceptibility but lack of
observable jump of the magnetization at the second (higher
pressure) ``meta-magnetic" phase transition from FM2 to FM3.
Unlike the picture for UGe$_2$, in UIr both transitions, FM1-FM2
and FM2-FM3, are far from the maximum $T_m(P_m)$, so in this case
one can hardly speculate that the {\it max} is produced by the
nearby jump of magnetization. UIr seems to be a U-type
spin-triplet ferromagnetic superconductor.

\section{Final remarks}

Finally, even in its simplified form, this theory has been shown
to be capable of accounting for a wide variety of experimental
behavior. A natural extension to the theory is to add a
$\mbox{\boldmath$M$}^6$ term which provides a formalism to
investigate possible meta-magnetic phase transitions
\cite{Huxley:2000} and extend some first order phase transition
lines. Another modification of this theory, with regard to
applications to other compounds, is to include a $P$ dependence
for some of the other GL parameters. The fluctuation and quantum
correlation effects can be considered by the respective
field-theoretical action of the system, where the order
parameters~$\mbox{\boldmath$\psi$}$ and $\mbox{\boldmath$M$}$ are
not uniform but rather space and time dependent. The vortex
(spatially non-uniform) phase due to the spontaneous magnetization
$\mbox{\boldmath$M$}$ is another phenomenon which can be
investigated by a generalization of the theory by considering
non-uniform order parameters fields $\mbox{\boldmath$\psi$}$ and
$\mbox{\boldmath$M$}$ (see, e.g., Ref.~\cite{Tewari:2004,
Li:2006}). Note, that such theoretical treatments are quite
complex and require a number of approximations. As already noted
in this paper the magnetic fluctuations stimulate first order
phase transitions for both finite and zero phase transition
temperatures.

{\bf ACKNOWLEDGEMENTS:} The authors thank M. G. Cottam for useful
discussions and A. Harada, and S. M. Hayden for valuable private
communications. One of us (D.I.U.) thanks the University of
Western Ontario for hospitality. Partial support from NFSR-Sofia
(through grant Ph. 1507) is also acknowledged.\


\begin{thebibliography}{ll}
\bibitem{Vollhardt:1990} D. Vollhardt and P. W\"olfle,
{it The Superfluid Phases of Helium 3}
(Taylor $\&$ Francis, London, 1990); D. I. Uzunov, in: {\em
Advances in Theoretical Physics}, edited by E. Caianiello (World
Scientific, Singapore, 1990), p. 96; M. Sigrist and K. Ueda, Rev.
Mod. Phys. {\bf 63}, 239 (1991).
\bibitem{Saxena:2000}
S. S. Saxena, P. Agarwal, K. Ahilan, F. M. Grosche, R. K. W.
Haselwimmer, M.J. Steiner, E. Pugh, I. R. Walker, S.R. Julian, P.
Monthoux, G. G. Lonzarich, A. Huxley. I. Sheikin, D. Braithwaite,
and J. Flouquet,   Nature {\bf 406}, 587 (2000).
\bibitem{Huxley:2001}
A. Huxley, I. Sheikin, E. Ressouche, N. Kernavanois, D.
Braithwaite, R. Calemczuk, and J. Flouquet,  Phys. Rev. {\bf B63},
144519 (2001).
\bibitem{Tateiwa:2001}
N. Tateiwa, T. C. Kobayashi, K. Hanazono, A. Amaya, Y. Haga. R.
Settai, and Y. Onuki, J. Phys. Condensed Matter {\bf 13}, L17
(2001).
\bibitem{Harada:2007}
A. Harada, S. Kawasaki, H. Mukuda, Y. Kitaoka, Y. Haga, E.
Yamamoto, Y. Onuki, K. M. Itoh, E. E. Haller, and H. harima, Phys.
Rev. B {\bf 75}, 140502 (2007).
\bibitem{Aoki:2001}
D. Aoki, A. Huxley, E. Ressouche, D. Braithwaite, J. Flouquet,
J-P.. Brison, E. Lhotel, and C. Paulsen,  Nature {\bf 413}, 613
(2001).
\bibitem{Hardy1:2005}
F. Hardy, A. Huxley, Phys. Rev. Lett. {\bf 94}, 247006 (2005).
\bibitem{Hardy2:2005}
F. Hardy, A. Huxley, J. Flouquet, B. Salce, G. Knebel, D.
Braithwate, D. Aoki, M. Uhlarz, and C. Pfleiderer, Physica B {\bf
359-361} 1111 (2005).
\bibitem{Huy:2007}
N. T. Huy, A. Gasparini, D. E. de Nijs, Y. Huang, J. C. P.
Klaasse, T. Gortenmulder, A. de Visser, A. Hamann, T. G\"orlach,
and H. v. L\"ohneysen, Phys. Rev. Lett. {\bf 99}, 067006 (2007).
\bibitem{Huy:2008}
N. T. Huy, D. E. de Nijs, Y. K. Huang, and A. de Visser, Phys.
Rev. Lett. {\bf 100}, 077001 (2008).
\bibitem{Akazawa:2005}
T. Akazawa, H. Hidaka, H. Kotegawa, T. C. Kobayashi, T. Fujiwara,
E. Yamamoto, Y. Haga, R. Settai, and Y. Onuki, Physica B {\bf
359-361}, 1138 (2005).
\bibitem{Kobayashi:2006} T. C. Kobayashi,S. Fukushima, H. Hidaka, H. Kotegawa,
 T. Akazawa, E. Yamamoto, Y. Haga, R. Settai, and Y. Onuki, Physica B {\bf 378-361},
378 (2006).
\bibitem{Pfleiderer:2001}
C. Pfleiderer, M. Uhlatz, S. M. Hayden, R. Vollmer, H. v.
L\"ohneysen, N. R. Berhoeft, and G. G. Lonzarich,  Nature {\bf
412}, 58 (2001).
\bibitem{Yelland1:2005} E. A. Yelland, S. J. C. Yates, O. Taylor, A. Griffiths, S. M.
Hayden, and A. Carrington, Phys. Rev. B {\bf 72}, 184436 (2005).
\bibitem{Yelland2:2005} E. A. Yelland, S. M. Hayden, S. J. C. Yates,
C. Pfleiderer, M. Uhlarz, R. Vollmer, H. v L\"ohneysen, N. R.
Bernhoeft, R. P. Smith, S. S. Saxena, and N. Kimura, Phys. Rev.
{\bf B72}, 214523 (2005).
\bibitem{Bolesh:2005} C. J. Bolesh and T. Giamarchi, Phys. Rev.
Lett. {\bf 71}, 024517 (2005); R. D. Duncan, C. Vaccarella, and C.
A. S. de Melo, Phys. Rev. B {\bf 64},  172503 (2001).
\bibitem{Nevidomskyy:2005} A. H. Nevidomskyy, Phys. Rev. Lett.
{\bf 94}, 097003 (2005).
\bibitem{Uzunov:1993}
D. I. Uzunov, {\em Theory of Critical Phenomena} (World
Scientific, Singapore, 1993).
\bibitem{Machida:2001}
K. Machida and T. Ohmi,  Phys. Rev. Lett. {\bf 86}, 850 (2001).
\bibitem{Walker:2002} M. B. Walker and K. V. Samokhin, Phys. Rev. Lett. {\bf 88},
207001 (2002); K. V. Samokhin and M. B. Walker, Phys. Rev. B {\bf
66}, 024512 (2002); Phys. Rev. B {\bf 66}, 174501 (2002).
\bibitem{Linder:2007}
J. Linder, A. Sudbo, Phys. Rev. B {\bf 76}, 054511 (2007); J.
Linder, I. B. Sperstad, A. H. Nevidomskyy, M. Cuoco, and A. Sodbo,
Phys. Rev. {\bf 77}, 184511 (2008); J. Linder, T. Yokoyama, and A.
Sudbo, Phys. Rev. B {\bf 78}, 064520 (2008); J. Linder, A. H.
Nevidomskyy, A. Sudbo, Phys. Rev. B {\bf 78}, 172502 (2008).
\bibitem{Shopova:2005}
D. V. Shopova and D. I. Uzunov, Phys. Rev. {\bf 72}, 024531
(2005); Phys. Lett. A {\bf 313}, 139 (2003).
\bibitem{Shopova:2006}
D. V. Shopova and D. I. Uzunov, in: {\em Progress in
Ferromagnetism Research}, ed. by V. N. Murray (Nova Science
Publishers, New York, 2006), p. 223;  D. V. Shopova and D. I.
Uzunov, J. Phys. Studies , {\bf 4}, 426 (2003) 426; D. V. Shopova
and D. I. Uzunov, Compt. Rend Acad. Bulg. Sci.
 {\bf 56}, 35 (2003) 35; D. V. Shopova, T. E. Tsvetkov, and D. I. Uzunov, Cond.
Matter Phys. {\bf 8}, 181 (2005) 181; D. V. Shopova, and D. I.
Uzunov, Bulg. J. of Phys. {\bf 32}, 81 (2005).
\bibitem{Cowley:1980}
R. A. Cowley,  Adv. Phys. {\bf 29}, 1 (1980); J-C. Tol\'edano and
P. Tol\'edano, {\em The Landau Theory of Phase Transitions} (World
Scientific, Singapore, 1987).
\bibitem{Yamada:1993}
K. K. Murata and S. Doniach, Phys. Rev. Lett. {\bf 29}, 285
(1972); G. G. Lonzarich and L. Taillefer, J. Phys. C: Solid State
Phys. {\bf 18}, 4339 (1985); T. Moriya, J. Phys. Soc. Japan {\bf
55}, 357 (1986); H. Yamada, Phys. Rev. B {\bf 47}, 11211 (1993).
\bibitem{Cottam:2008}
M. G. Cottam, D. V. Shopova and D. I. Uzunov, Phys. Lett. A
(2008), in press.
\bibitem{Shopova:2003}
D. V. Shopova and D. I. Uzunov, Phys. Rep. C {\bf 379}, 1 (2003).
\bibitem{Wohlfarth:1968} E. P. Wohlfarth, J. Appl. Phys. {\bf 39},
1061 (1968); Physica B$\&$C {\bf 91B}, 305 (1977).
\bibitem{Misra:2008} P. Misra, {\it Heavy-Fermion Systems},
(Elsevier, Amsterdam, 2008).
\bibitem{Sandeman:2003} K. G. Sandeman, G. G. Lonzarich, and A. J.
Schofield, Phys. Rev. Lett. {\bf 90}, 167005 (2003).
\bibitem{Smith:1971} T. F. Smith, J. A. Mydosh, and E. P.
Wohlfarth, Phys. rev. Lett. {\bf 27}, 1732 (1971); G. Oomi, T.
Kagayama, K. Nishimura, S. W. Yun, and Y. Onuki, Physica B {\bf
206}, 515 (1995).
\bibitem{Kimura:2004}
N. Kimura \textit{et al.}, Phys. Rev. Lett. {\bf 92 }, 197002
(2004).
\bibitem{Uhlarz:2004}
M. Uhlarz, C. Pfleiderer, and S. M. Hayden, Phys. Rev. Lett. {\bf
93}, 256404 (2004).
\bibitem{Uzunov:2006}
D. I. Uzunov, Phys. Rev. {\bf B74}, 134514 (2006); Europhys. Lett.
{\bf 77}, 20008 (2007).
\bibitem{Belitz:2005} D. Belitz, T. R. Kirkpatrick, J.
Rollb\"uhler, Phys. Rev. Lett. {\bf 94}, 247205 (2005); G. A.
Gehring, Europhys. Lett. {\bf 82}, 60004 (2008).
\bibitem{Halperin:1974}
 B. I. Halperin, T. C. Lubensky, and S. K. Ma, Phys. Rev. Lett.
{\bf 32}, 292 (1974); J-H. Chen, T. C. Lubensky, and D. R. Nelson,
{\em Phys. Rev.} {\bf B17}, 4274 (1978).
\bibitem{Huxley:2000}
A. Huxley, I. Sheikin, and D. Braithwaite, Physica {\bf B
284-288}, 1277 (2000).
\bibitem{Tewari:2004} S Tewari, D. Belitz, T. R. Kirkpatrick, and
J. Toner, Phys. Rev. Lett. {\bf 93}, 177002 (2004).
\bibitem{Li:2006} Q. Li, D. Belitz, and T. R. Kirkpatrick, Phys.
Rev. B {\bf 74}, 134505 (2006).
\end{thebibliography}
\end{document}